# NONLINEAR EVOLUTION OF DENSITY PERTURBATIONS*

J.S. Bagla[1] and T. Padmanabhan[2]
Inter-University Centre for Astronomy and Astrophysics
Post Bag 4, Ganeshkhind,
Pune - 411 007, INDIA.

## Abstract

From the epoch of recombination ($z \approx 10^3$) till today, the typical density contrasts have grown by a factor of about $10^6$ in a Friedmann universe with $\Omega = 1$. However, during the same epoch the typical gravitational potential has grown only by a factor of order unity. We present theoretical arguments explaining the origin of this approximate constancy of gravitational potential. This fact can be exploited to provide a new, powerful, approximation scheme to study the formation of nonlinear structures in the universe. The essential idea of this method is to evolve the initial distribution of particles using a gravitational potential frozen in time. We carry out this scheme for several standard models including the CDM and HDM and show that the results match quite well with those obtained by exact numerical simulations. We compute different statistical measures of clustering and compare them for the description of nonlinear evolution.

This approximation also provides valuable insight into understanding various features of nonlinear evolution; for example, it provides a simple explanation as to why pancakes remain thin during the evolution even in the absence of any artificial, adhesion-like, damping terms. We also compare this approximation with other schemes like Zeldovich approximation and frozen-flow. Our procedure has a far greater range of validity than the Zeldovich approximation since it can handle motion across (and inside) a caustic properly. Unlike in frozen-flow, actual shell-crossing does occur in the frozen-potential approximation; hence it provides a far more accurate description of the velocity field compared to frozen flow approximation.

## 1. Introduction

It is believed that structures like galaxies and clusters of galaxies formed out of small inhomogeneities via gravitational instability. In this paper we discuss a new approximation scheme called Frozen Potential Approximation (hereafter FPA) ( Bagla and Padmanabhan 1993a ; Brainerd et. al. 1993) for studying the nonlinear growth of the inhomogeneities. In section 2 we introduce and compare the various approximation schemes and in section 3 we discuss the results obtained from numerical simulations. In section 4 we present some analytical results which can be obtained using FPA, and the limits of applicability of FPA are studied in section 5.

## 2. Quasilinear Approximations

In studying the process of structure formation, linear perturbation theory can be used when the density contrast $\delta$ is small but one has to rely on full N-body simulations to grasp the nonlinear evolution. Our understanding of processes in the nonlinear regime is very limited due to the absence of analytical results for the nonlinear phase. However, there exist some approximate schemes for evolving the trajectories of particles which are valid in the quasi-linear phase. These approximations are very useful in understanding the physical processes in the nonlinear regime.

The exact trajectory of a particle in a Friedmann universe is described by the equations

$$\frac{d^2\mathbf{x}}{dt^2} + 2\frac{\dot{a}}{a}\frac{d\mathbf{x}}{dt} = -\frac{1}{a^2}\nabla\phi; \quad \nabla^2\phi = 4\pi G \varrho_b a^2 \delta \qquad (1)$$

where $a(t)$ is the expansion factor, $\mathbf{x}$ is the comoving coordinate, and $\rho_b$ is the background matter density. We shall set $\Omega = 1$ and consider perturbations only in the matter dominated era, though the results can be easily generalised for other cases. These equations can be recast using scale factor as the time variable, and by introducing the variable $\psi = (2/3H_o^2)\phi$. In terms of $a$ and $\psi$, (1) and (2) become

$$\frac{d^2\mathbf{x}}{da^2} + \frac{3}{2a}\frac{d\mathbf{x}}{da} = -\frac{3}{2a}\nabla\psi; \quad \nabla^2\psi = \frac{\delta}{a} \qquad (2)$$

As we shall see, various approximations follow very naturally from these equations.

It is apparent from (2) that the velocity of particles is not constant but decays even when $\nabla \psi = 0$ due to the damping term $(3/2a)(d\mathbf{x}/da)$ in (2). In Zeldovich approximation (hereafter ZA) ( Zeldovich 1970 ; Shandarin and Zeldovich 1989), it is assumed that the velocity of a particle remains constant along its trajectory and has the value it had at the starting point. This is equivalant to taking

$$\frac{d^2\mathbf{x}}{da^2} = 0; \quad \frac{d\mathbf{x}}{da} = -\nabla\psi(\mathbf{x}_{in})$$

so that the trajectory in ZA is

$$\mathbf{x} = \mathbf{x}_{in} - a\nabla\psi|_{in} \qquad (3)$$

By assuming a uniform velocity, in contrast to the decaying velocity in the absence of a potential, some amount of accleration has been taken into account. This approximation works well till the trajectories of particles cross but leads to unphysical results after crossing. To see how this happens, consider the motion of particles near a potential well. All the velocity vectors in the neighbourhood of a minima of the potential point towards it and all the particles move towards the minima. However, after the particles cross the minima, they do not turn around and fall back into the minima, as they should in a self gravitating system of particles. Instead, the particles continue travelling with their initial velocity leading to dispersal of structures. Hence this approximation works well only if the average displacement of particles is smaller than the scale over which $\nabla\psi$ changes sign.

Zeldovich ansatz can be used to show that first structures to form in a generic gravitational collapse are planar surfaces of high density, the so called pancakes. The unrealistic part of dynamics in ZA is that particles after collapsing and forming a pancake continue to move with the same velocity leading to thickening of pancakes.

The Frozen flow approximation (FFA) ( Matarrese et. al. 1992 ) overcomes the thickening of pancakes by assuming that the motion of particles is well approximated by potential flow. In FFA the inertia of particles is assumed to play a negligible role in comparison with the expansion of the Universe or a rapidly ( spatially) varying potential. Therefore the first term in (2) can be neglected and the motion may be described by

$$\frac{d^2\mathbf{x}}{da^2} = 0; \quad \frac{d\mathbf{x}}{da}(\mathbf{x},a) = -\nabla\psi(\mathbf{x},a_{in}) \qquad (4)$$

The gravitational potential at the initial time generates a velocity field which is used to move the particles. Particles reach the minima of potential only asymptotically and the pancakes do not thicken, since (strictly speaking) no pancake has really formed. The dynamics in this approximation is unrealistic near pancakes as it does not predict oscillations of particles about the pancakes, a generic feature of motion in such a case.

It is possible to generate a better approximation (Bagla and Padmanabhan, 1993a; Brainerd et al., 1993) by the following procedure : we know that the gravitational potential is a constant in the linear regime for an $\Omega = 1$ Universe. It can be argued ( based on models like spherical top hat) that the potential does not change significantly even in the nonlinear regime. This approximate constancy of potential can be used to evolve density perturbations to mildly non-linear density contrasts with much more realistic dynamics. Frozen Potential Approximation (FPA) *assumes* that the potential remains constant even in the nonlinear regime. The trajectories are then evolved using (2) but with a fixed potential:

$$\frac{d^2\mathbf{x}}{da^2} + \frac{3}{2a}\frac{d\mathbf{x}}{da} = -\frac{3}{2a}\nabla\psi(\mathbf{x},a_{in}) \qquad (5)$$

By using the full nonlinear equation, a realistic dynamical evolution of trajectories is obtained. In FPA, near a minima, the particles move towards the minima of the potential and then oscillate about it with decaying amplitude due to the expansion of the Universe. In this approximation, shell crossing does occur and the pancakes are formed; the particles oscillate about the pancake with a small amplitude and move along the pancakes towards the minima of the potential.

The trajectories of a pair of particles in a potential well is plotted for the three approximations ( ZA, FFA, FPA ) in Fig.1. The potential is taken to be $\psi = -\cos(x)$ and the trajectory of particles is plotted on the $x - a$ plane. In Zeldovich approximation, the particles move with uniform velocity and even after reaching the minima of potential, continue travelling with the same velocity, erasing the structure in the process; the trajectory in this case is therefore a straight line ( dot – dashed ). In FFA, the particles approach the minima only asymptotically ( dashed line ), giving the impression of a stable structure even though no structure has really formed in finite time. On the other hand, FPA gives trajectories oscillating about the minima with a decaying amplitude ( unbroken line ). In exact N-body simulations the damping is more prominent and the pancakes are thinner than the ones obtained by FPA, because FPA does not take into account the deepening of potential wells as particles cluster.

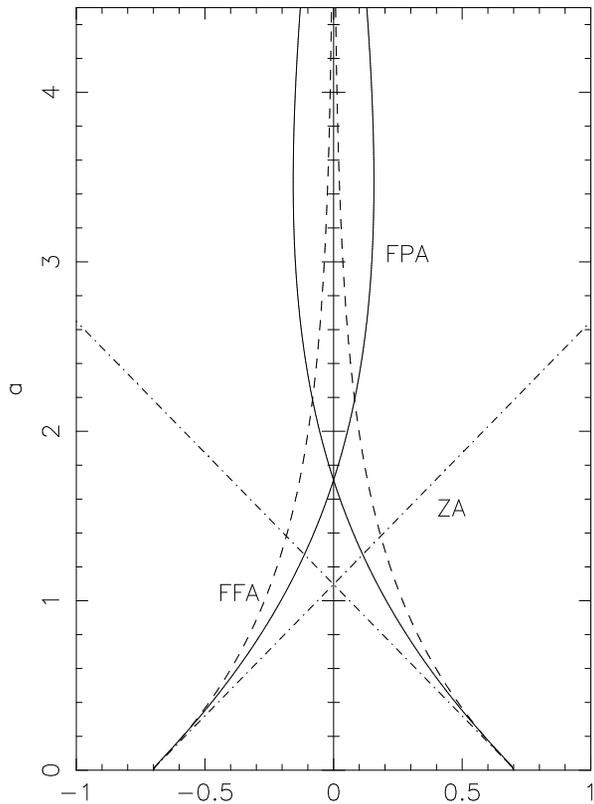

*Fig.1 : Trajectories of particles in a one dimensional potential well in various approximations are shown in the $x - a$ plane. The dot–dashed line corresponds to ZA, the dashed line to FFA and the continuous line to FPA. Notice that in FPA, the particles cross the potential minima and oscillate about it.*

## 3. Numerical Simulations

We tested three approximations using numerical simulations for models such as the CDM and HDM and compared the results with exact N-body simulations. In the simulations, we start with an initial potential $\psi$ which is a realisation of a gaussian random field with a power spectrum corresponding to the model under consideration $P_\psi(k) = P_\delta(k)/k^4$. The power spectra were normalised by using the COBE quadrapole value. The potential was obtained on the lattice sites of a cubic $(128 \times 128 \times 128)$ grid. The particles were moved by the ansatzs described above.

Various approximations can be compared visually in Fig.2 and Fig.3 which give slices [ with dimensions $(96 \times 96 \times 15)$ $(h^{-1}\mathrm{MPc})^3$ ] of the Universe in these schemes at different epochs for CDM and HDM respectivly. It is obvious that at early stages ($z \gg 1$) of structure formation, all approximations give similar results, while at later stages ($z \simeq 0$) the differences are quite prominent. Zeldovich approximation is the worst of the three; after shell crossing, the particles continue moving with the same initial velocity leading to thickening of the pancakes. In FFA the particles are deposited in the pancakes and then they move slowly towards the minima; the pancakes are very thin and there is very little clumpiness even for CDM spectrum. Pancakes do not thicken in FPA and, at late stages, clumpy structures dominate.

In Fig.4 we have shown the density contrast $\sigma(r)$ defined as $\sigma^2 \equiv \overline{\xi}$ with

$$\overline{\xi}(r) = \frac{3}{r^3} \int_0^r \xi(x) dx^3 = \frac{3J_3(r)}{r^3}$$

for three redshifts for standard CDM using FPA. For comparison we have also plotted the N-body result at $z = 0$. We have obtained the N-body result by using the universal scaling relation between the linear density contrast and the actual nonlinear density contrast (Hamilton et.al. 1991 ; Nityananda and Padmanabhan 1993). In the linear regime the evolution is self similar, i.e. the density contrast grows in proportion with the scale factor. When high density contrast is reached, FPA tends to underestimate the density contrast.

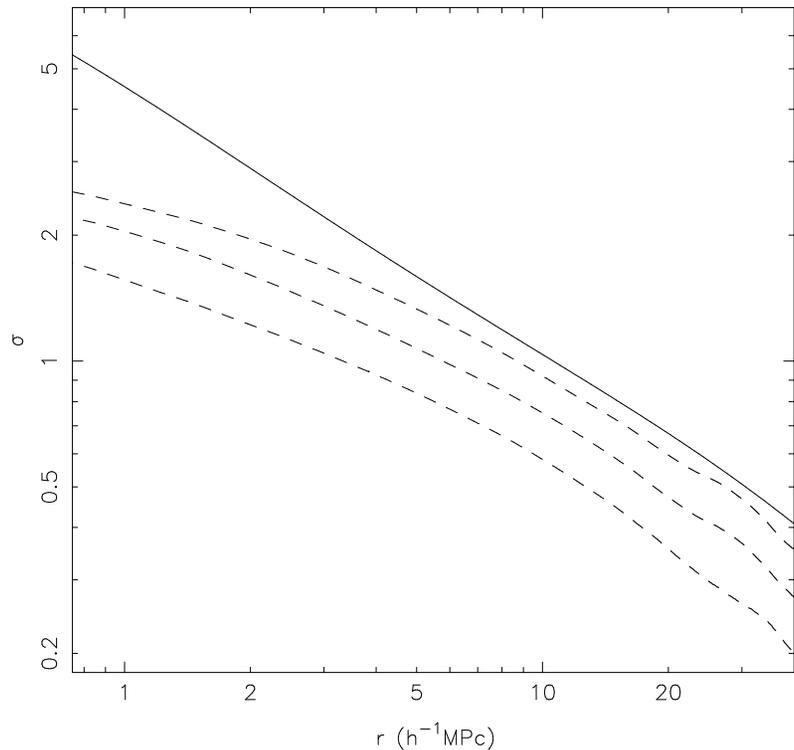

$Fig.4$ : *Evolution of density contrast in the Frozen Potential Approximation. Also drawn here is the curve for density contrast obtained from N-Body for $z = 0$. Here the density contrast is defined as $\sigma \equiv \sqrt{\bar{\xi}}$.*

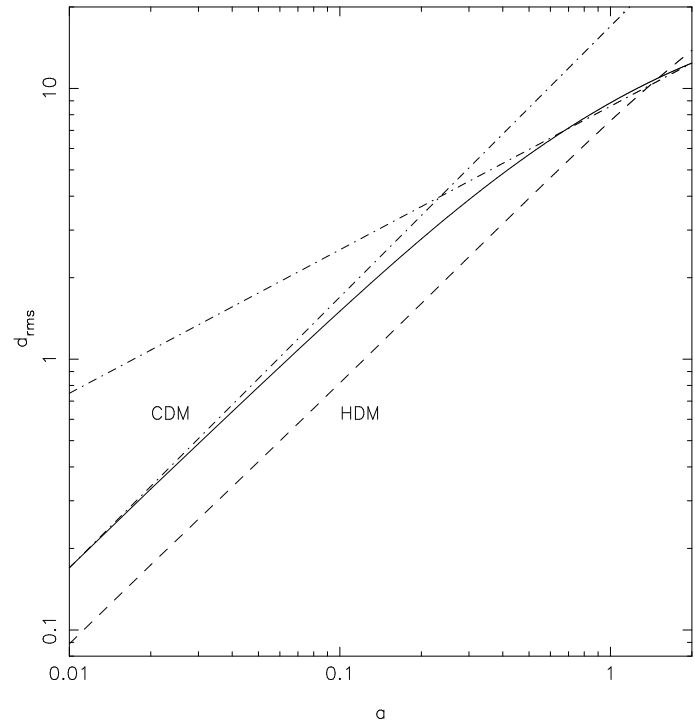

$Fig.5$ : *Evolution of $d_{rms}$. Here $d_{rms}$ is plotted as a function of the scale factor, the continuous line is for CDM and the dashed line corresponds to HDM model. Dot-dashed lines have been drawn for the power law solutions, $a$ and $a^{1/1.8}$ for early and late times respectively, as discussed in the text.*

The FPA can be quite useful in studying the evolution of voids in the Universe. To compare the sizes of voids with theoretical results it is necessary to choose a suitable definition of void. One way of estimating the sizes of voids is to compute the rms displacement of particles in a simulation (Shandarin, 1992) which can be related to the average diameter of voids by $D_{void} = 2k d_{rms}$ where $k$ is a constant of order unity which depends on the shape of voids. For spherical voids $k = 1.3$. The thick line in Fig.5 shows the evolution of $d_{rms}$ for CDM and HDM. For the standard CDM model, we found that $d_{rms} \simeq 18 h_{50}^{-1}$ MPc at $z = 0$.

## 4. FPA : Analytical approach

The equation (5) can be solved analytically in some special cases (Bagla and Padmanabhan, 1993b). Consider, for example, the motion of particles near a local density minima. The density contrast very close to the extrema is a constant and can be represented by $\delta = -\Delta$ with $0 < \Delta \le 1$ a constant. Since for constant density, $\nabla \psi = -r\Delta/3$, (5) becomes

$$\frac{d^2 r}{da^2} + \frac{3}{2a}\frac{dr}{da} = \frac{\Delta}{2a} r \qquad (6)$$

This has the solution

$$r = \frac{r_0}{\sqrt{2\Delta a}}\sinh(\sqrt{2\Delta a}).$$

Since the density contrast is valid for small $r$ and hence for small $a$; in this limit, $r \propto a$, as expected from linear theory. This suggests that the radius of voids increases linearly with the scale factor for $a \ll 1$.

The density profile away from the centre of the void can be approximated as

$$\delta = -\left(\frac{r}{L}\right)^{-n} \qquad (7)$$

where $n$ is a positive constant. This leads to the equation

$$\frac{d^2r}{da^2} + \frac{3}{2a}\frac{dr}{da} = \frac{3}{2a}\frac{L^n}{(3-n)}r^{1-n} \qquad (8)$$

This nonlinear equation has one simple solution :

$$r = \left[\frac{3n^2}{(3-n)(n+2)}\right]^{1/n} L a^{1/n}. \qquad (9)$$

which holds for $n < 3$.

For a gaussian random field, the profile of density around a density extrema is a power law with the same index as the correlation function( Bardeen et. al. 1986 (BBKS)). For our Universe, the index of the correlation function is about 1.8, therefore we expect the voids to grow as $a^{1/1.8}$ or $(1+z)^{-1/1.8}$ in the late stages. In the linear limit, we expect the void radius to grow as $a$. While calculating $d_{rms}$ from simulations, we average over regions with different index $n$ and hence expect $d_{rms}$ to grow at a rate intermediate between $a$ and $a^{1/1.8}$. At late times merging of voids will slow down the growth rate even further.

These features are shown in fig 5. The two dot-dash lines indicate growth proportional to $a$ and $a^{1/1.8}$ and have intercepts chosen to match with $d_{rms}$ for CDM. It is clear that these lines match with $d_{rms}$ at early and late times respectively. Since the degree of nonlinearity is lower in HDM compared to CDM (when normalised to COBE), the $d_{rms} \propto a$ for HDM in this range.

Similar analysis can be carried out for generic motion near a density peaks and it can be shown that very close to the peak, the solutions are oscillatory.

## 5. FPA : Accuracy and Limitations

The results above show that FPA is a good approximation for a CDM spectrum and other similar spectra. The FPA is based on the assumption that the potential $\psi$ changes very little during the evolution of density perturbations. To understand the agreement with N-body simulations and to study the limitations of FPA, we should ask the question: how much does the gravitational potential change in reality ?

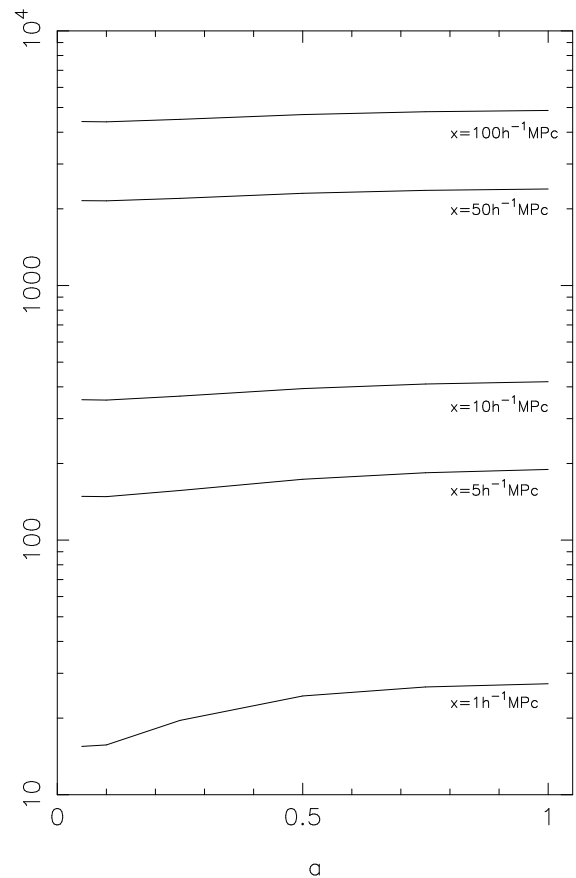

*Fig.6 : Variation in Potential with scale factor. We have plotted the variance of potential $\zeta$ as a funtion of the scale factor for $r = 1, 5, 10, 50$ and $100 h^{-1}$ MPc.*

At large scales, at which $\sigma^2 \lesssim 1$ and linear theory is valid, the gravitational potential is a constant to a high degree of accuracy. Similarly at very small scales – dominated by fully virialised structures – the gravitational potential does not change significantly. The evolution of gravitational potential in the quasi-linear regime depends on the power spectrum. It is possible to obtain a semianalytic form for this evolution by using the form of the scaled relative pair velocity $h \equiv (v_{pair}/Hr)$ where $v_{pair}(x)$ is the average relative pair velocity at the scale $x$ and $r = ax$. From the N-body simulations, it has been observed that $h(\mathbf{x},t) = h[\sigma^2(x,t)]$ is a universal function of $\sigma^2$ (Hamilton et al., 1991; Nityananda and Padmanabhan, 1993). We can use this universal relation to obtain a relation between the linear density contrast and the nonlinear density contrast. We find that the relation between true density contrast $\sigma^2$ and the linear density contrast $\sigma_L^2$ is well approximated by:

$$\sigma^2(x,a) \propto \begin{cases} \sigma_L^2(l,a) & \text{(for } \sigma^2 \lesssim 1 \text{)} \\ [\sigma_L^2(l,a)]^3 & \text{(for } 3 \lesssim \sigma^2 \lesssim 50) \\ [\sigma_L^2(l,a)] & \text{( for } 50 \lesssim \sigma^2) \end{cases} \qquad (10)$$

where $l = x(1+\sigma^2)^{1/3}$ (Bagla and Padmanabhan, 1993b). If we approximate $\sigma_L^2(l,a) \propto a^2 l^{-(n+3)}$ where $n$ is a local index, this gives us for $(\sigma/a)$:

$$\frac{\sigma(x,a)}{a} \propto \begin{cases} a^{-\frac{(n+1)}{(n+4)}} & 10 \lesssim \sigma^2 \lesssim 50 \\ a^{-\frac{(n+2)}{(n+5)}} & 50 \lesssim \sigma^2 \end{cases} \qquad (11)$$

In the standard CDM model $n \simeq -2$ in the nonlinear domain ($\sigma^2 \gtrsim 50$) and $n \simeq (-1 \text{ to } -2)$ in the quasilinear regime. From (11) we see that $(\sigma/a)$ remains constant for $n = -1$ and $n = -2$ in the quasilinear and nonlinear regimes respectivly. Hence in CDM like spectra, there is a conspiracy of indices ensuring that there is very little change in the gravitational potential. The above analysis, of course, is approximate since the relation between $\psi$ and $(\sigma/a)$ is nonlocal and the index $n$ varies with scale. The true variation of $\psi$ can be obtained by integrating the Poisson's equation numerically. The Fig 6 shows $\psi$ at four different scales in standard CDM model, as a function of redshift.

The FPA, however, is not accurate at small scales, for a different reason. The location of pancake like structures in our ansatz is influenced significantly by the initial potential. This implies that the large scale structures hardly move in our ansatz. In reality if the streaming velocity at a (large) length scale $L$ is $v(L)$ then structures at this scale would have, on the average moved a distance of $H_0^{-1} v(L)$. In CDM models at scales $L \simeq 50 h^{-1} Mpc$, $v \simeq 200 km s^{-1}$; hence these structures could have moved about $H_0^{-1} v \simeq 2 h^{-1} Mpc$ by now. Thus at scales smaller than about $1 h^{-1} Mpc$ we would expect the approximation to show some inaccuracy.

These aspects can be clearly seen in fig. 7a, b which compares FPA with exact N-body simulations in 2D. The pancakes in N-body simulations are thinner than those in FPA (due to deepening of potential wells) and are also slightly shifted with respect to those in FPA.

## 6. Discussion and Conclusions

The FPA is a powerful approximation which can be used to obtain approximate results, numerically as well analytically. It compares well with the N-body simulations in a statistical sense. By construction, the minima of potential do not move in this approximation and this makes it inaccurate at small scales. A very strong point in favour of this approximation is that it contains full information about velocities and therefore it can be used to study various models in the redshift space. The FPA can also be used to study the motion of baryons in dark matter potential wells.

**Figure Captions** (For figures not included here)

**Fig.2** : Evolution of density perturbations in CDM for various approximations. These frames, from left to right, correspond to $a = 0.25, 0.5$ and $1.0$ ( present epoch) respectively. The top row of frames is for frozen potential approximation, the middle shows evolution according to the frozen flow approximation and the bottom row is for Zeldovich approximation.

**Fig.3** : Same as Fig.2 but for HDM at $a = 0.5, 1.0$ and $2.0$.

**Fig.7a** : An FPA simulation in 2-d. Here we have used a power law spectrum with $n = -1$.

**Fig.7b** : An N-body frame from a 2-d simulation for the same spectrum as in Fig.7a. The pancakes are much thinner than the ones in the FPA simulation. Also notice the shift of pancakes in this case.

*THESE FIGURES ARE AVAILABLE  
FROM THE AUTHORS ON REQUEST*